# Upper Critical Fields up to 60T in Dirty Magnesium Diboride Thin Films

C. Ferdeghini, V. Ferrando, C. Tarantini, E. Bellingeri, G. Grasso, A. Malagoli, D. Marrè, M. Putti,
P. Manfrinetti, A. Pogrebnyakov, J.M. Redwing, X.X. Xi, R. Felici and E. Haanappel

*Abstract*—Upper critical fields of several magnesium diboride thin films were measured up to 28 T at the Grenoble High Magnetic Field Laboratory (GHMFL) in Grenoble and up to 60 T at the Laboratoire National des Champs Magnétiques Pulsés (LNCMP) in Toulouse. The samples were prepared both by pulsed laser deposition (PLD) and hybrid physical chemical vapour deposition (HPCVD) technique; they have critical temperatures between 29 and 39 K and normal state resistivities between 5 and 250 μΩ cm; one of them has been intentionally doped with carbon. The measured critical fields, $\mu_0 H_{c2}$, were exceptionally high; we obtained the record value of 52 T at 4.2 K in the parallel orientation. In contrast with the BCS predictions, no saturation in $H_{c2}$ at low temperature was observed. Furthermore, films with a wide range of resistivity values showed similar critical fields, suggesting that in a two band system resistivity and $H_{c2}$ are not trivially linked. The high $H_{c2}$ values seem to be related with the expanded c-axis. The structure of one of the samples was carefully investigated with X-ray diffraction at European Synchrotron Radiation Facility (ESRF) in Grenoble.

## I. INTRODUCTION

SINCE the discovery of its superconductivity [1], magnesium diboride appeared as a charming material, not only for the high critical temperature, but also for the several unusual properties arising from its two band nature. In fact, the two separate sheets of Fermi surface [2], [3] give rise to two distinct energy gaps. The larger gap $\Delta_\sigma$ is associated with the anisotropic and nearly two-dimensional σ bands, while the smaller gap $\Delta_\pi$ is associated to the isotropic π bands. The σ

Manuscript received October 5, 2004. The work at Penn State was supported in part by ONR under grant Nos. N00014-00-1-0294 (X.X.X.) and N0014-01-1-0006 (J.M.R.).

C. Ferdeghini, C. Tarantini, E. Bellingeri, G. Grasso, A. Malagoli, D. Marrè, M. Putti, and P. Manfrinetti are with the LAMIA/INFM c/o Dipartimento di Fisica, via Dodecaneso 33, 16146 Genova, Italy (phone: +390103536282; fax: +39311066; e-mail: ferdeghini@ fisica.unige.it).

V. Ferrando is with the Pennsylvania State University, University Park, PA, USA, on leave from the LAMIA/INFM, Dipartimento di Fisica, via Dodecaneso 33, 16146 Genova, Italy.

A. Pogrebnyakov, J.M. Redwing, and X.X. Xi are with the Pennsylvania State University, University Park, PA, USA.

R. Felici is with INFM-OGG – ESRF 6, rue J. Horowitz, BP220 38043 Grenoble, France.

E. Haanappel is with the Laboratoire National des Champs Magnétiques Pulsés, CNRS-UPS-INSA, Toulouse, France.

and π bands can be regarded as different channels conducting in parallel giving that the interband impurity scattering is negligible with respect to the intraband ones. It is only within this complex scenario that a number of superconducting properties can be explained [4], [5], [6]. In particular, the effect of two bands on critical field behavior is now intensively studied both theoretically and experimentally. The upper critical field in $MgB_2$ is anisotropic with the critical field parallel to the *ab* planes ($H_{c2}^\parallel$) higher than the perpendicular one ($H_{c2}^\perp$). Single crystals, in which the clean limit condition is easily reached, show relatively low critical fields ($\mu_0 H_{c2}^\perp$ = 3-5 T and $\mu_0 H_{c2}^\parallel$ = 16-19 T) [7], [8], [9]. High purity polycrystals presented critical fields in the same 16-19T range [10]. Also thin films in the clean limit, such as those deposited by the HPCVD technique [11], show critical fields similar to those of single crystals. It is well known that defects in the structure, in a single band superconductor, increase the normal state resistivity and, proportionally, the critical field value. This has been verified for low temperature superconductors and represents the usual method to enhance $H_{c2}$ in technological materials, such as Nb-Ti and A15 compounds.

The same approach does not work trivially for magnesium diboride: in fact, here, the multiband nature of the superconductivity has to be taken into account. Recently, theoretical articles describing the critical field behaviour in this framework appeared in the literature both for the dirty limit [12], [13], [14], and for the clean limit [15]. Focusing the attention on the dirty limit, the model proposed by Gurevich [12] considers the intra-band electronic diffusivities $D_\pi$ and $D_\sigma$ (the inverse of scattering rates), neglecting the inter-band ones. At low (high) temperatures the upper critical field is determined by the smaller (larger) diffusivity. The resistivity, instead, is determined by the two conducting channels in parallel and thus is dominated by the higher diffusivity. Therefore the proportionality between normal state resistivity and critical field value in $MgB_2$ can be lost. The shape of $H_{c2}(T)$ can be considerably different from the BCS one and $H_{c2}(0)$ can drastically exceed the BCS prediction. There is a wide literature devoted to the increase of the critical fields by the introduction of defects starting from clean $MgB_2$ in single crystal or polycrystalline form. These defects were introduced by neutron (or proton and heavy ions) irradiation [16], by varying the Mg stoichiometry with subsequent annealing [17], and by doping with C in the B plane [18], [19] or with Al in Mg planes [20].

In all bulk materials when the resistivity increases the critical field is also increased by a factor 2 (up to 30-35T ) except for the case of Al substitution in which the critical field is depressed: this fact was explained, in a clean limit scenario, by the depression of $\Delta_\sigma$ [20]. The case of thin films is different in fact, higher critical fields, up to 60T, near the paramagnetic limit, can be reached [21], [22], [23].

Moreover, Ferrando et al. [21] showed that there is no dependence of critical field values on $\rho$ in a set of films with $\rho$ varying by one order of magnitude.

In this paper we present high magnetic field measurements performed in different experiments in 23T and 28T resistive magnets at Grenoble High Magnetic Field Laboratory, France and in 60 T pulsed magnet at Laboratoire National des Champs Magnétiques Pulsés, Toulouse, France on a set of $MgB_2$ thin films. We present also structural characterization of one of these sample performed with synchrotron radiation at ESFR, Grenoble, France.

## II. EXPERIMENTAL AND DISCUSSION

### A. The samples

The samples considered here (for which the critical field was measured at $\mu_0 H > 20T$) are eight different films. Seven of them (Film 1-7) were prepared at INFM-LAMIA of Genova. They were grown on different substrates by pulsed laser ablation from a stoichiometric target followed by annealing in a Mg atmosphere in a tantalum tube at 900 °C for 30 minutes, as described in detail elsewhere [24]. Film 1 is an epitaxial film on c-cut $Al_2O_3$ [25] and shows $T_C = 29.5$ K and $\rho_{40K} = 40$ $\mu\Omega$cm. Film 2, Film 3 and Film 5 are c-oriented thin films grown on MgO(111) and present $T_C$ =32.0, 33.9 and 33.5K and $\rho_{40K}$ =50, 20 and 47$\mu\Omega$cm respectively. Film 4 and Film 6 were grown on c-cut $Al_2O_3$ and were determined to be, using x-rays analysis, c-oriented with $T_C$= 38.8 and 33.3K and $\rho_{40K}$= 5 and 43$\mu\Omega$cm respectively. The data for Film 1-4 have already been reported in [21]. Film 7 is an epitaxial film deposited on $Al_2O_3$ c-cut with a zirconium diboride buffer layer [26] with $T_C$=37.6 K and $\rho(40)$ of about 12$\mu\Omega$cm. Film 8 was grown at Penn State University by HPCVD [11] with the addition of 75sccm of $(C_5H_5)_2Mg$ to the $H_2$ carrier gas. It show $T_C$= 35.0K and a huge value of 250$\mu\Omega$cm as normal state resistivity.

The properties of all films are summarized in Table 1.

### B. Critical field measurements

Measurements of $H_{c2}(T)$ on Film 1-4, Film 6 and Film 7 were performed in resistive magnets (23 and 28T) at GHMFL in Grenoble. $R(H)$ curves were measured in the parallel and perpendicular orientations at a sweep rate of 1 T/min with a standard four probe AC technique, while temperature was stabilized to about 10 mK. We used a current density in the range 10-100 A/cm$^2$. Measurements on Film 5 and Film 8 were performed in the 300 msec 60T pulsed facility at LNCMP in Toulouse, using a high-frequency ($f$ = 40 kHz) digital lock-in technique. The sample current density was in the range 50-200 A/cm$^2$ with no change in the $R(H)$ curves. In all the cases $H_{c2}$ was defined as the 90% of the resistive transition.

TABLE I
SAMPLES CHARACTERISTICS

| FILM | $T_c$, K | $\Delta T_c$, K | $\rho$, $\mu\Omega$cm | RRR | Substrate | c, Å |
|---|---|---|---|---|---|---|
| 1 | 29.5 | 2 | 40 | 1.2 | $Al_2O_3$c-cut | 3.517[a] |
| 2 | 32.0 | 1.5 | 50 | 1.3 | MgO(111) | 3.532[a] |
| 3 | 33.9 | 1.1 | 20 | 1.5 | MgO(111) | 3.533[a] |
| 4 | 38.8 | 1 | 5 | 2.5 | $Al_2O_3$c-cut | 3.519[a] |
| 5 | 33.5 | 1 | 47 | 1.5 | MgO(111) | 3.533[a] |
| 6 | 33.3 | 0.8 | 43 | 1.3 | $Al_2O_3$c-cut | 3.523[a] |
|   |      |     |    |     |                | 3.524[b] |
|   |      |     |    |     |                | 3.558[b] |
| 7 | 37.6 | 1 | 12 | 1.8 | $Al_2O_3$c-cut +$ZrB_2$ | 3.545[a] |
| 8 | 35.1 | 0.8 | 250 | 1.1 | SiC(001) | 3.536[a] |

[a] from X-Rays Diffraction analyses.
[b] from Synchrotron Radiation Diffraction analyses.

Fig. 1 shows the upper critical field curves as a function of temperature for all the eight samples in the parallel (to the ab planes) (upper panel) and perpendicular (lower panel) orientations.

We first discuss the critical field values in the parallel orientation. Here, a possibly incomplete c-orientation does not affect the $H_{c2}^\parallel$ values. For Film 8 we measured the record value of 52 T at 4.2 K. Film 8 is the film grown by HPCVD with carbon doping: the C-doping increases the film resistivity from about 1 $\mu\Omega$cm, the typical value for an undoped sample [11], to 250 $\mu\Omega$cm. It is worth noticing that this critical field value is greater than the critical fields presented by bulk C-doped samples with the same critical temperature [18]. Although Film 8, the film with the highest resistivity, shows the highest critical field, all the films present critical fields higher than those of defective bulk samples. In particular Film 2, 3, 4, 7 present similar $H_{c2}$ values, at least down to the minimum temperature for which the measurement were possible using the resistive magnet at GHMFL. It is worth noting that the normal state resistivity of Film 4 is 5 $\mu\Omega$cm, 50 times less than that of Film 8. The independence of the critical field on the normal state resistivity, claimed in [21] is strongly supported. This is in contrast with the standard BCS theory, for which the upper critical field $H_{c2}(0)$ is given by: $H_{c2}(0)=0.69 \cdot T_C \cdot (dH_{c2}/dT)$, where $dH_{c2}/dT$ is proportional to the normal state residual resistivity. All the samples, in the perpendicular orientation, and Film 5 and 8, in parallel one, have been measured at fields high enough to allow a linear extrapolation and a reasonable estimation of $H_{c2}(0)$. These data highlight a linear temperature dependence and any saturation at low temperature; this behavior demonstrates that the BCS prediction strongly underestimates the experimental values.

In the inset to Fig. 1, the critical field anisotropies $\gamma$ of the eight samples are reported. The low temperature γ values range from 2.0 (Film 4) to 3.5 (Film 2) and γ decreases with increasing temperature. These values are considerably lower with respect to the clean single crystal ones. In general, in $MgB_2$, disorder decreases γ as clearly shown in C-doped single crystals [27].

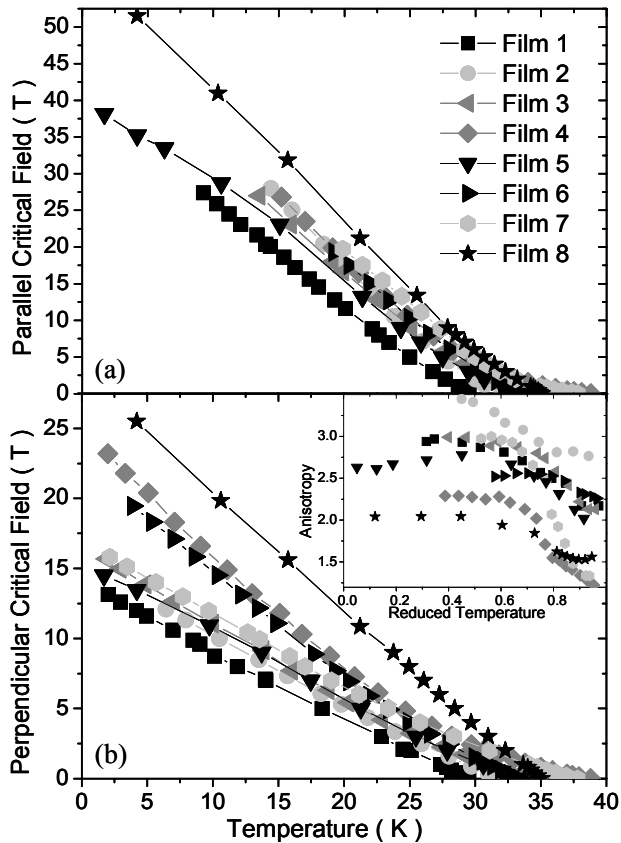

Fig. 1. Critical fields as a function of temperature for parallel (upper panel) and perpendicular (lower panel) orientations. In the inset: the anisotropy factor behavior.

With the data of Fig. 1 we have the confirmation that critical fields of thin films are considerably higher than those of bulk materials with any type of defect. Recently, Braccini *et al.* [23] presented extrapolated critical field values in C-doped dirty thin film approaching the paramagnetic limit of 70 T. Considering a set of 12 films coming from various laboratories, they noted how the samples that show high critical fields, have an expanded *c*-axis. They verified with TEM analyses on one of these films [22] the presence of buckling of the Mg planes that naturally causes *c*-axis lattice expansion; the lattice buckling induces strong π scattering and, therefore high critical fields. They hypothesized that a similar mechanism can work also for the other films that presented high critical fields. In spite of the fact that this hypothesis is convincing considering the set of samples considered, we present here at least three films, Film 1, Film 4 and Film 6, that apparently show, according to the XRD analyses, *c*-axis smaller than the bulk value. They are all deposited on *c*-cut sapphire. For this reason we decided to carefully characterize the structure of this kind of films using the powerful tool of synchrotron radiation diffraction.

### C. X-Ray diffraction at a synchrotron radiation facility

Film 6, among the films on $Al_2O_3$, was characterized by synchrotron radiation diffraction at ESFR ID01 in Grenoble. The film is 160 nm thick and was discovered not to be single phase. This sample presents an MgO epitaxial layer (8 nm) between the substrate and the superconductor, probably formed during the high temperature annealing in Mg atmosphere. Moreover we identified two structurally distinct $MgB_2$ phases.

In Fig. 2 a reciprocal space map (*HK* plane) of the region around the 102 reflection of $MgB_2$ is shown. Three different peaks are present, one due to MgO and the two others attributed to $MgB_2$ with slightly different in plane *a*-cell parameter. Performing cuts of the reciprocal space in the *L* direction, the *c*-axes of these two phases were obtained as shown in Fig. 3 where *L* scans on the A and B maxima of the *HK*-map are reported in the right and left panels respectively. The two $MgB_2$ phases differ both in the *c* axis value and in crystallographic order as is evident in the different peak broadening. The phase with the higher crystalline quality (B) shows lattice parameters close to the bulk ones ($a_B$=3.083 Å and $c_B$=3.524 Å) whereas the other has an expanded cell ($a_A$=3.094 Å and $c_A$=3.558Å) and a largely strained (≈2%) structure. The large broadening of the peak A and the low scattering power of $MgB_2$, can explain why the double phase was not detected with standard laboratory x-ray diffraction experiments. The values of the cell parameters, determined by synchrotron measurements, could be affected by systematic errors, so that the absolute values can be inaccurate. However, since the lattice parameters of the more intense peak are close to the ones measured by conventional x-ray techniques, we are confident in the correct estimation of the other values.

The analyses of the X-ray diffraction data at ESRF showed that also in the films grown on sapphire (like Film 6), there is at least a part of the sample with expanded *c*-axis. Therefore the hypothesis that the *c*-axis enhancement and the high critical fields are correlated is further supported.

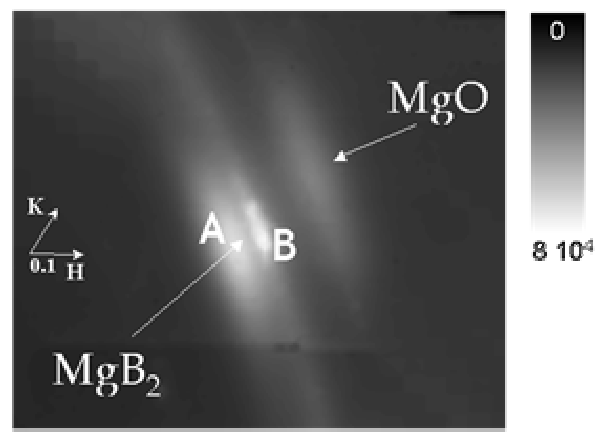

Fig. 2. HK reciprocal space map around $MgB_2$ 102 reflection. Two different $MgB_2$ phases are identified. A peak of the MgO epitaxial layer is also present.

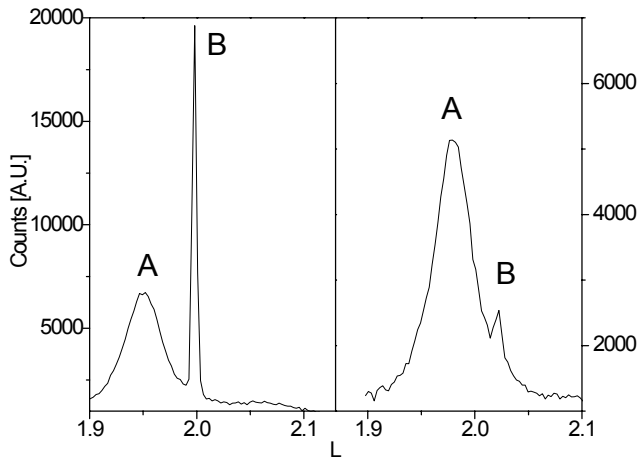

Fig. 3. Reciprocal space L scans performed on the B (left panel) and A (right panel) maximum of the map in Fig 2.

## III. Conclusion

In this paper we showed the upper critical fields of some Magnesium Diboride thin films measured at fields higher than 20 T. The measured critical fields are exceptionally high; we obtained the record value of 52 T at 4.2 K in the parallel orientation in a high resistivity C-doped sample. The $H_{c2}$ versus temperature behavior is not compatible with the BCS predictions, i.e. no saturation in $H_{c2}$ being observed at low temperature. Furthermore, the measured films showed similar critical fields in spite of the wide range of resistivity values, indicating that resistivity and $H_{c2}$ are not trivially linked. The high $H_{c2}$ values in these films seem to be related to the expanded *c*-axis as confirmed by structural characterization with X-ray diffraction at ESRF.

## References

[1] J.Nagamatsu, N.Nakawaga, T.Muranaka, Y.Zenitani, J.Akimitsu, "Superconductivity at 39 K in magnesium diboride," *Nature,* vol.410 pp.63-65, 2001.
[2] A.Y.Liu, I.I.Mazin , J.Kortuss, "Beyond Eliashberg Superconductivity in MgB$_2$: Anharmonicity, Two-Phonon Scattering, and Multiple Gaps," *Phys.Rev.Lett,* vol. 87 pp.087005-9, 2001.
[3] H.J.Choi, D. Roundy, H. Sun, M. L. Cohen, S. G. Louie, "First-principles calculation of the superconducting transition in MgB$_2$ within the anisotropic Eliashberg formalism," *Phys. Rev. B,* vol.66, pp.020513, 2002.
[4] F.Bouquet et al., **"**Phenomenological two-gap model for the specific heat of MgB$_2$," *Europhys. Lett.*, Vol.56, pp.856-862, 2001.
[5] A.Brinkman et al. "Multiband model for tunneling in MgB$_2$ junctions," *Phys. Rev. B,* vol.65, pp.180517(R), 2002.
[6] M.Putti et al., " Thermal conductivity of MgB$_2$ in the superconducting state. " *Phys. Rev. B,* vol.67, pp.064505, 2003.
[7] L.Lyard et al. "Anisotropy of the upper critical field and critical current in single crystal MgB$_2$," *Phys. Rev. B,* vol. 66, pp.180502, 2002.
[8] M. Angst et al., "Temperature and Field Dependence of the Anisotropy of MgB$_2$," *Phys. Rev. Lett,*. Vol. 88 , pp.167004, 2002.
[9] A.V.Sologubenko, J.Jun , S.M.Kazakov, J.Karpinski, H.R.Ott, "Temperature dependence and anisotropy of the bulk upper critical field H$_{c2}$ of MgB$_2$," *Phys. Rev. B,* vol.65, pp.180505, 2002.
[10] M.Putti et al., "Critical fields of MgB$_2$: crossover from clean to dirty limit," *Phys. Rev. B,* vol.70, pp.052509,. 2004.
[11] X. Zeng et al., "*In situ* epitaxial MgB$_2$ thin films for superconducting electronics," *Nat. Mater.,* Vol.1, pp.35-38, 2002.
[12] A.Gurevich, "Enhancement of the upper critical field by nonmagnetic impurities in dirty two-gap superconductors," *Phys. Rev. B,* vol. 67, pp. 184515, 2003.
[13] T.Dahm, N.Schopohl, "Fermi surface topology and the upper critical field in two-band superconductors: Application to MgB$_2$," *Phis. Rev. Lett.*, vol.91, pp.017001, 2003.
[14] A.A. Golubov, A.E.Koshelev , "Upper critical field in dirty two-band superconductors: Breakdown of the anisotropic Ginzburg-Landau theory," *Phys. Rev .B,* vol. 68, pp. 104503, 2003.
[15] P.Miranovic, M.Machida, V.G.Cogan, "Anisotropy of the upper critical field in superconductors with anisotropic gaps: Anisotropy parameters of MgB$_2$," *J. Phys. Soc. Jpn* , vol.72, pp.221-224, 2003.
[16] M.Putti et al.,"Neutron irradiation of Mg$^{11}$B$_2$: from the enhancement to the suppression of superconducting properties,",*Appl. Phys. Lett.*, submitted for publication.
[17] Braccini et al., "Significant enhancement of irreversibility field in clean-limit bulk MgB$_2$," *Appl. Phys. Lett.*, Vol.81, pp.4578-4579, 2002.
[18] R.H.T.Wilke et al., "Systematic Effects of Carbon Doping on the Superconducting Properties of Mg(B$_{1-x}$C$_x$)$_2$," *Phys. Rev. Lett.,* Vol.92 pp.217003, 2004.
[19] E.Ohmichi T.Masui, S.Lee, S.Tajima, T.Osada, "Enhancement of the irreversibility field by carbon substitution in single crystal MgB$_2$," *cond-mat/0312348*
[20] M.Putti et al., "Critical field of Al-doped MgB$_2$ samples: correlations with the suppression of the σ gap," *Phys. Rev. Lett,*. submitted for publication.
[21] V.Ferrando et al., "Effect of two bands on critical fields in MgB$_2$ thin films with various resistivity values," *Phys. Rev. B*, vol.68, pp. 094517, 2003.
[22] A.Gurevich et al., "Very high upper critical fields in MgB$_2$ produced by selective tuning of impurity scattering," *Supercond. Sci. Technol.,* vol.17 pp. 278-286, 2004.
[23] Braccini et al., "High-field superconductivity in alloyed MgB$_2$ thin films," *Cond-mat/0402001*
[24] C.Ferdeghini et al., "Growth of *c*-oriented MgB$_2$ thin films by pulsed laser deposition: structural characterization and electronic anisotropy," *Supercond. Sci. Technol.,* Vol.14 pp.952-957, 2001.
[25] V.Ferrando et al., "Growth methods of c-axis oriented MgB$_2$ thin films by pulsed laser deposition," *Supercond. Sci. Technol.*, vol.16, pp.241-245, 2003.
[26] V.Ferrando et al., "Epitaxial MgB$_2$ thin films on ZrB$_2$ buffer layers: structural characterization by synchrotron radiation," *Supercond. Sci. Technol.*, in press.
[27] T.Masui, S.Lee, S.Tajima, "Carbon-substitution effect on the electronic properties of MgB$_2$ single crystal," *Phys. Rev. B,* vol.70, pp.024504, 2004.